\documentclass{llncs}
\usepackage{amssymb}
\usepackage{epsfig}
\usepackage{makeidx}  

\newenvironment{keywords}{
       \list{}{\advance\topsep by0.35cm\relax\small
       \leftmargin=1cm
       \labelwidth=0.35cm
       \listparindent=0.35cm
       \itemindent\listparindent
       \rightmargin\leftmargin}\item[\hskip\labelsep
                                     \bfseries Keywords:]}
     {\endlist}

\begin{document}

\begin{frontmatter}

\pagestyle{headings}  

\mainmatter              

\title{Complex Systems with Trivial Dynamics}

\titlerunning{Complex Systems with Trivial Dynamics}  

\author{Ricardo L\'{o}pez-Ruiz}

\authorrunning{Ricardo L\'{o}pez-Ruiz}   

\institute{Dept. of Computer Science, Faculty of Science and Bifi, \\ 
Universidad de Zaragoza, 50009 - Zaragoza, Spain,\\
\email{rilopez@unizar.es}}

\maketitle              

\begin{abstract} 
In this communication, complex systems with a near trivial
dynamics are addressed. First,
under the hypothesis of equiprobability in the asymptotic equilibrium,
it is shown that the (hyper) planar geometry of an $N$-dimensional
multi-agent economic system implies the exponential
(Boltzmann-Gibss) wealth distribution and that the spherical geometry
of a gas of particles implies the Gaussian (Maxwellian) distribution
of velocities. Moreover, two non-linear models are proposed to
explain the decay of these statistical systems from an
out-of-equilibrium situation toward their asymptotic equilibrium states.
\end{abstract}


\end{frontmatter}

\section{Introduction}
\label{S1} 

In this paper, different classical results \cite{huang1987,yako2009} are recalled.
They are obtained from a  geometrical interpretation of different multi-agent systems 
evolving in phase space under the hypothesis of equiprobability  \cite{lopez2007,lopez2008}. 
Two nonlinear models that explain the decay of these statistical systems to
their asymptotic equilibrium states are also collected \cite{lopez2011,shivanian2012}.

We sketch in section \ref{S2} the derivation of the Boltzmann-Gibbs (exponential) distribution \cite{lopez2008}
by means of the geometrical properties of the volume of an $N$-dimensional pyramid.
The same result is obtained when the calculation is performed over the surface of 
a such $N$-dimensional body. In both cases, the motivation is a multi-agent economic system
with an open or closed economy, respectively.  

Also, a continuous version of an homogeneous economic gas-like 
model \cite{lopez2011} is given in the section \ref{S2}. This model explains the appearance,
independently of the initial wealth distribution given to the system,
of the exponential (Boltzmann-Gibbs) distribution as the asymptotic equilibrium in random markets,
and in general in many other natural phenomena with the same type of interactions.

The Maxwellian (Gaussian) distribution is derived in section \ref{S3}
from geometrical arguments over the volume or the surface of an $N$-sphere \cite{lopez2007}.
Here, the motivation is a multi-particle gas system in contact
with a heat reservoir (non-isolated or open system) or with a fixed energy
(isolated or closed system), respectively. 

The ubiquity of the Maxwellian velocity distribution 
in ideal gases is also explained in the section \ref{S3} 
with a nonlinear mapping acting in the space 
of velocity distributions \cite{shivanian2012}. This mapping is an operator 
that gives account of the decay of any initial velocity 
distribution toward the Gaussian (Maxwellian) distribution.

Last section contains the conclusions.

\section{Systems Showing the Boltzmann-Gibbs Distribution}
\label{S2}

\subsection{Multi-Agent Economic Open Systems}

Here we assume $N$ agents, each one with coordinate $x_i$, $i=1,\ldots,N$, 
with $x_i\geq 0$ representing the wealth or money of the agent $i$,
and a total available amount of money $E$:
\begin{equation}
x_1+x_2+\cdots +x_{N-1}+x_N \leq E.
\label{eq-e}
\end{equation} 
Under random or deterministic evolution rules for the exchanging of money among agents,
let us suppose that this system evolves in the interior of the $N$-dimensional pyramid 
given by Eq. (\ref{eq-e}). The role of a heat reservoir, that in this model
supplies money instead of energy, could be played by the state or by the bank system 
in western societies.  
The formula for the volume $V_N(E)$ of an equilateral $N$-dimensional pyramid 
formed by $N+1$ vertices linked by $N$ perpendicular sides of length $E$ is
\begin{equation}
V_N(E) = {E^N\over N!}.
\label{eq-S_n1}
\end{equation}
We suppose that each point on the $N$-dimensional pyramid is equiprobable, 
then the probability $f(x_i)dx_i$ of finding 
the agent $i$ with money $x_i$ is proportional to the 
volume formed by all the points into the $(N-1)$-dimensional pyramid 
having the $i$th-coordinate equal to $x_i$. 
Then, it can be shown that the Boltzmann factor
(or the Maxwell-Boltzmann distribution), $f(x_i)$, is given by
\begin{equation}
f(x_i) = {V_{N-1}(E-x_i)\over V_N(E)},
\label{eq-f_n1}
\end{equation}
that verifies the normalization condition
\begin{equation}
\int_{0}^Ef(x_i)dx_i = 1.
\label{eq-p_n1}
\end{equation}
The final form of $f(x)$, in the asymptotic regime $N\rightarrow\infty$ 
(which implies $E\rightarrow\infty$) and taking the mean wealth $\epsilon=E/N$, is: 
\begin{equation}
f(x)dx = {1\over \epsilon}\,e^{-{x/\epsilon}}dx,
\label{eq-gauss11}
\end{equation}
where the index $i$ has been removed because the distribution is the same for each agent, 
and thus the wealth distribution can be obtained by averaging over all the agents. 
This distribution has been found to fit the real distribution of incomes 
in western societies \cite{yako2001}.

\subsection{Multi-Agent Economic Closed Systems}

We derive now the Boltzmann-Gibbs distribution by considering the system 
in isolation, that is, a closed economy. Without loss of generality, 
let us assume $N$ interacting economic agents, each one with coordinate 
$x_i$, $i=1,\ldots,N$, with $x_i\geq 0$, and where $x_i$ represents an amount of money. 
If we suppose that the total amount of money $E$ is conserved,
\begin{equation}
x_1+x_2+\cdots +x_{N-1}+x_N = E,
\label{eq-E}
\end{equation} 
then this isolated system 
evolves on the positive part of an equilateral $N$-hyperplane. 
The surface area $S_N(E)$ of an equilateral 
$N$-hyperplane of side $E$ is given by
\begin{equation}
S_N(E) = {\sqrt{N}\over (N-1)!}\;E^{N-1}.
\label{eq-S_n2}
\end{equation}
If the ergodic hypothesis is 
assumed, each point on the $N$-hyperplane is equiprobable. 
Then the probability $f(x_i)dx_i$ of finding 
agent $i$ with money $x_i$ is proportional to the 
surface area formed by all the points on the $N$-hyperplane having the $i$th-coordinate 
equal to $x_i$. It can be shown that $f(x_i)$ is the Boltzmann factor (Boltzmann-Gibbs distribution),
with the normalization condition (\ref{eq-p_n1}).
It takes the form,
\begin{equation}
f(x_i) = {1\over S_N(E)}
{S_{N-1}(E-x_i)\over \sin\theta_N},
\label{eq-f_n}
\end{equation}
where the coordinate $\theta_N$ satisfies 
$\sin\theta_N = \sqrt{N-1 \over N}$.
After some calculation the Boltzmann distribution 
is newly recovered:
\begin{equation}
f(x)dx = {1\over k\tau}\,e^{-x/k\tau}dx,
\end{equation}
with $\epsilon=k\tau$, being $k$ the Boltzmann constant and 
$\tau$ the temperature of the statistical system.

\subsection{The Continuous Economic Gas-like Model}

We consider an ensemble of economic agents trading with money in a random manner \cite{yako2001}.
This is one of the simplest gas-like models, in which an initial amount of money is given
to each agent, let us suppose the same to each one. Then, pairs of agents are randomly chosen
and they exchange their money also in a random way. When the gas evolves under these conditions,
the exponential distribution appears as the asymptotic wealth distribution. In this model,
the microdynamics is conservative because the local interactions conserve
the money. Hence, the macrodynamics is  also conservative and the total amount of money is
constant in time.

The discrete version of this model is as follows \cite{yako2001}.
The trading rules for each interacting pair $(m_i,m_j)$ of the ensemble of $N$ economic
agents can be written as
\begin{eqnarray}
m'_i &=& \sigma \; (m_i + m_j), \nonumber\\
m'_j &=& (1 - \sigma) (m_i + m_j), \label{model1}\\
i , j &=& 1 \ldots N, \nonumber
\end{eqnarray}
where $\sigma$ is a random number in the interval $(0,1)$.
The agents $(i,j)$ are randomly chosen. Their initial money $(m_i, m_j)$,
at time $t$, is transformed after the interaction in $(m'_i, m'_j)$ at time $t+1$.
The asymptotic distribution $p_f(m)$, obtained by numerical simulations,
is the exponential (Boltzmann-Gibbs) distribution,
\begin{equation}
p_f(m)=\beta \exp(-\beta \,m), \hspace{0.5cm}\hbox{with}\hspace{0.5cm}\beta={1/ <m>_{gas}},
\label{eq-exp}
\end{equation}
where $p_f(m) {\mathrm d}m$ denotes the PDF ({\it probability density function}), i.e.
the probability of finding an agent with money (or energy in a gas system) between
$m$ and $m + {\mathrm d}m$.
Evidently, this PDF is normalized, $\vert\vert p_f\vert\vert=\int_0^{\infty} p_f(m){\mathrm d}m=1$.
The mean value of the wealth, $<m>_{gas}$, can be easily calculated directly from the gas
by $<m>_{gas}=\sum_i m_i/N$.

The continuous version of this model \cite{lopezruiz2011} considers the evolution of
an initial wealth distribution $p_0(m)$ at each time step $n$ under the action of an operator $T$.
Thus, the system evolves from time $n$ to time $n+1$ to asymptotically
reach the equilibrium distribution $p_f(m)$, i.e.
\begin{equation}
\lim_{n\rightarrow\infty} {T}^n \left(p_0(m)\right) \rightarrow p_f(m).
\label{eq-operT}
\end{equation}
In this particular case, $p_f(m)$ is the exponential distribution with the same
average value $<p_f>$ than the initial one $<p_0>$,
due to the local and total richness conservation.

The derivation of the operator $T$ is as follows \cite{lopezruiz2011}.
Suppose that $p_n$ is the wealth distribution in the ensemble at time $n$.
The probability to have a quantity of money $x$ at time $n+1$ will be the sum of the
probabilities of all those pairs of agents $(u,v)$ able to
produce the quantity $x$ after their interaction, that is, all the pairs verifying $u+v>x$.
Thus, the probability that two of these agents with money $(u,v)$ interact between them is
$p_n(u)*p_n(v)$. Their exchange is totally random and then they can give rise with equal
probability to any value $x$ comprised in the interval $(0,u+v)$. Therefore, the probability
to obtain a particular $x$ (with $x<u+v$) for the interacting pair $(u,v)$ will be
$p_n(u)*p_n(v)/(u+v)$.  Then, $T$ has the form of a nonlinear integral operator,
\begin{equation}
p_{n+1}(x)={T}p_n(x) = \int\!\!\int_{u+v>x}\,{p_n(u)p_n(v)\over u+v}
\; {\mathrm d}u{\mathrm d}v \,.
\label{eq-T}
\end{equation}

If we suppose $T$ acting in the PDFs space, it has been proved \cite{lopez2011}
that $T$ conserves the mean wealth of the system, $<Tp>=<p>$. It also conserves
the norm ($\vert\vert \cdot\vert\vert$), i.e. $T$ maintains the total number of agents
of the system, $\vert\vert T p\vert\vert=\vert\vert p\vert\vert=1$, that
by extension implies the conservation of the total richness of the system.
We have also shown that the exponential distribution $p_f(x)$ with the right average value
is the only steady state of $T$, i.e. $T p_f=p_f$. Computations also seem to suggest
that other high period orbits do not exist.
In consequence, it can be argued that the relation (\ref{eq-operT}) is true.
This decaying behavior toward the exponential distribution 
is essentially maintained in the extension of this model for more general random markets.

\section{Systems Showing the Maxwellian Distribution}
\label{S3}

\subsection{Multi-particle open systems}

Let us suppose a one-dimensional ideal gas of $N$ non-identical 
classical particles with masses $m_i$, with $i=1,\ldots,N$, and total 
maximum energy $E$. If particle
$i$ has a momentum $m_iv_i$, we define a kinetic energy:
\begin{equation}
K_i \equiv p_i^2 \equiv {1 \over 2}{ m_iv_i^2},
\label{eq-p_i}
\end{equation} 
where $p_i$ is the square root of the kinetic energy $K_i$. 
If the total maximum energy is defined as $E \equiv R^2$, we have 
\begin{equation}
p_1^2+p_2^2+\cdots +p_{N-1}^2+p_N^2 \leq R^2.
\label{eq-Ee}
\end{equation} 
We see that the system has accessible states with different energy, which can be 
supplied by a heat reservoir. These states are all those enclosed into the volume 
of the $N$-sphere given by Eq. (\ref{eq-Ee}). 
The formula for the volume $V_N(R)$
of an $N$-sphere of radius $R$ is
\begin{equation}
V_N(R) = {\pi^{N\over 2}\over \Gamma({N\over 2}+1)}R^{N},
\label{eq-S_n3}
\end{equation}
where $\Gamma(\cdot)$ is the gamma function. If we suppose that each point
into the $N$-sphere is equiprobable, then the probability $f(p_i)dp_i$ of finding 
the particle $i$ with coordinate $p_i$ (energy $p_i^2$) is proportional to the 
volume formed by all the points on the $N$-sphere having the $i$th-coordinate 
equal to $p_i$. It can be shown that
\begin{equation}
f(p_i) = {V_{N-1}(\sqrt{R^2-p_i^2})\over V_N(R)},
\label{eq-f_n2}
\end{equation}
which is normalized, $\int_{-R}^Rf(p_i)dp_i = 1$. The Maxwellian distribution 
is obtained in the asymptotic regime $N\rightarrow\infty$ 
(which implies $E\rightarrow\infty$):
\begin{equation}
f(p)dp = \sqrt{1\over 2\pi\epsilon}\,e^{-{p^2/2\epsilon}}dp,
\label{eq-gauss}
\end{equation}
with $\epsilon=E/N$ being the mean energy per particle and where the index $i$ 
has been removed because the distribution is the same for each particle. 
Then the equilibrium velocity distribution can also be obtained by averaging
over all the particles.

\subsection{Multi-particle closed systems}

We start by assuming a one-dimensional ideal gas of $N$ non-identical 
classical particles with masses $m_i$, with $i=1,\ldots,N$, and total 
energy $E$. If particle $i$ has a momentum $m_iv_i$, newly we define a 
kinetic energy $K_i$ given by Eq. (\ref{eq-p_i}), where $p_i$ is the square 
root of $K_i$. If the total energy is defined as $E \equiv R^2$, we have 
\begin{equation}
p_1^2+p_2^2+\cdots +p_{N-1}^2+p_N^2 = R^2.
\label{eq-E1}
\end{equation} 
We see that the isolated system evolves on the surface of an $N$-sphere. 
The formula for the surface area $S_N(R)$
of an $N$-sphere of radius $R$ is
\begin{equation}
S_N(R) = {2\pi^{N\over 2}\over \Gamma({N\over 2})}R^{N-1},
\label{eq-S_n4}
\end{equation}
where $\Gamma(\cdot)$ is the gamma function. If the ergodic hypothesis is 
assumed, that is, each point on the $N$-sphere is equiprobable, 
then the probability $f(p_i)dp_i$ of finding 
the particle $i$ with coordinate $p_i$ (energy $p_i^2$) is proportional to the 
surface area formed by all the points on the $N$-sphere having the $i$th-coordinate 
equal to $p_i$. It can be shown that 
\begin{equation}
f(p_i) = {1\over S_N(R)}
{S_{N-1}(\sqrt{R^2-p_i^2})\over (1-{p_i^2\over R^2})^{1/2}},
\label{eq-f_n3}
\end{equation}
which is normalized. Replacing  $p^2$ by ${1\over 2}mv^2$,
$f(p)$ takes the following form $g(v)$ in the asymptotic limit $N\rightarrow\infty$,
\begin{equation}
g(v)dv = \sqrt{m\over 2\pi k\tau}\,e^{-{mv^2/2k\tau}}dv.
\end{equation}
This is the typical form of the Maxwellian distribution, 
with $\epsilon=k\tau/2$ given by the equipartition theorem.

\subsection{The Continuous Model for Ideal Gases}

Here, as we have done in the anterior case of economic systems,
we present a new model to explain the Maxwellian distribution as a limit point
in the space of velocity distributions for a gas system evolving from any initial 
condition \cite{shivanian2012}. 

Consider an ideal gas with particles of unity mass in the three-dimensional ($3D$) space. 
As long as there is not a privileged direction in the equilibrium, we can take any direction
in the space and study the discrete time evolution of the velocity distribution in that direction.
Let us call this direction $U$. We can complete a Cartesian system with two additional orthogonal 
directions $V,W$. If $p_n(u){\mathrm d}u$ represents the probability of finding 
a particle of the gas with velocity component in the direction $U$ comprised between 
$u$ and $u + {\mathrm d}u$ at time $n$, then the probability to have at this time $n$
a particle with a $3D$ velocity $(u,v,w)$ will be $p_n(u)p_n(v)p_n(w)$.
 
The particles of the gas collide between them, and after a number of interactions
of the order of system size, a new velocity distribution is attained at time $n+1$. 
Concerning the interaction of particles with the bulk
of the gas, we make two simplistic and realistic assumptions in order to obtain
the probability of having a velocity $x$ in the direction $U$ at time $n+1$:
(1) Only those particles with an energy bigger than $x^2$ at time $n$ can contribute 
to this velocity $x$ in the direction $U$, that is, all those particles whose velocities 
$(u,v,w)$ verify $u^2+v^2+w^2\ge x^2$; (2) The new velocities after collisions are equally 
distributed in their permitted ranges, that is, 
particles with velocity $(u,v,w)$ can generate maximal velocities
$\pm U_{max}=\pm\sqrt{u^2+v^2+w^2}$, then the allowed range of velocities $[-U_{max},U_{max}]$
measures $2|U_{max}|$, and the contributing probability of these particles to the velocity $x$
will be $p_n(u)p_n(v)p_n(w)/(2|U_{max}|)$. Taking all together we finally get the expression 
for the evolution operator $T$. This is: 
\begin{equation}
p_{n+1}(x)=Tp_n(x) = \int\int\int_{u^2+v^2+w^2\ge x^2}\,{p_n(u)p_n(v)p_n(w)\over 2\sqrt{u^2+v^2+w^2}}
\; {\mathrm d}u{\mathrm d}v{\mathrm d}w\,.
\end{equation}

Let us remark that we have not made any supposition about the type of interactions or collisions
between the particles and, in some way, the equivalent of the Boltzmann hypothesis of {\it molecular
chaos} \cite{boltzmann} would be the two simplistic assumptions we have stated on the interaction 
of particles with the bulk of the gas. Then, an alternative framework
than those usually presented in the literature \cite{maxwell} appears now on the scene. 
In fact, it is possible to show that the operator $T$ conserves in time
the energy and the null momentum of the gas. Moreover, for any initial velocity
distribution, the system tends towards its equilibrium, i.e. towards the Maxwellian
velocity distribution (1D case). This means that
\begin{equation}
\lim_{n\rightarrow\infty} T^n \left(p_0(x)\right) \rightarrow 
p_{\alpha}(x)=\sqrt{\alpha\over\pi}e^{-\alpha x^2}
\end{equation}
with $\alpha=(2\,<x^2,p_0>)^{-1}$.
In physical terms, it means that for any initial velocity distribution of the gas, 
it decays to the Maxwellian distribution, which is just the fixed point of the dynamics.
Recalling that in the equilibrium $<x^2,p_{\alpha}>=k\tau$, with $k$ the Boltzmann constant 
and $\tau$ the temperature of the gas, and introducing the mass $m$ of the particles, 
let us observe that the Maxwellian velocity distribution can be recovered in its $3D$ format: 
\begin{equation}
p_{\alpha}(u)p_{\alpha}(v)p_{\alpha}(w)=\left(m\alpha\over\pi\right)^{3\over 2}\,
e^{-m\alpha (u^2+v^2+w^2)} \hskip 0.5cm with \hskip 5mm \alpha=(2k\tau)^{-1}.
\end{equation}
In general, it is observed that the convergence of the $T$-iterations of any distribution 
$p(x)$ to its Gaussian limit $p_{\alpha}(x)$ is very fast.

\section{Conclusion}
\label{S4} 

We have shown that the Boltzmann factor describes 
the general statistical behavior of each small part 
of a multi-component system whose components or parts are given
by a set of random variables that satisfy an additive constraint,
in the form of a conservation law (closed systems) or 
in the form of an upper limit (open systems).

Let us remark that these calculations do not need the knowledge of the exact or microscopic 
randomization mechanisms of the multi-agent system in order to reach the equiprobability.
In some cases, it can be reached by random forces, in other cases 
by chaotic or deterministic causes.
Evidently, the proof that these mechanisms generate equiprobability is not a trivial task
and it remains as a typical challenge in this kind of problems.

In order to explain the ubiquity and stability of this type of distributions two models
based on discrete mappings in the space of distributions have been proposed.
On one hand, the gas-like models interpret economic exchanges of money between agents
similarly to collisions in a gas where particles share their energy.
The continuous version of a gas-like discrete model where the agents trade 
in binary collisions has been introduced to explain the stability of the exponential distribution
in this kind of economic systems.
On the other hand, a nonlinear map acting on the velocity distribution space of ideal gases,
which gives account of the decay of an out-of-equilibrium velocity distribution
toward the Maxwellian distribution, has been presented. Some properties 
concerning the dynamical behavior of both operators have also been sketched. 

 \subsection*{Acknowledgements}
 Several collaborators have participated in the development of different aspects 
 of this line of research. Concretely, X. Calbet, J. Sa\~nudo, J.L. Lopez and E. Shivanian.
 See the references.

\end{document}